\documentclass[fleqn,twoside]{article}
\usepackage{espcrc2}

\usepackage{graphicx}
\usepackage[figuresright]{rotating}

\newcommand{\beq}{\begin{eqnarray}}
\newcommand{\eeq}{\end{eqnarray}}
\newcommand{\CL}{{\cal L}}
\newcommand{\CD}{{\cal D}}
\newcommand{\Tr}{{\rm Tr\,}}
\newcommand\eqn[1]{\label{eq:#1}} 
\newcommand\Eq[1]{Eq.~(\ref{eq:#1})} 
\newcommand{\sla}[1]%
        {\kern .25em\raise.18ex\hbox{$/$}\kern-.75em #1}
\newcommand{\myeq}%
        {\kern-.75em &=& \kern-.75em}
\newcommand{\myplus}%
        {\kern-.75em &+& \kern-.75em}
\newcommand{\myminus}%
        {\kern-.75em &-& \kern-.75em}
\newcommand{\mybar}[1]%
        {\kern 0.8pt\overline{\kern -0.8pt#1\kern -0.8pt}\kern 0.8pt}

\newcommand{\AmS}{{\protect\the\textfont2
  A\kern-.1667em\lower.5ex\hbox{M}\kern-.125emS}}

\title{Lattice Supersymmetry}

\author{David B. Kaplan\address[INT]{Institute for Nuclear Theory\\ 
        Box 351550\\University of Washington\\Seattle, WA 98195-1550, USA}}%

\begin{document}

\begin{abstract}
A method is proposed for latticizing a class of supersymmetric gauge
theories, including N=4 super Yang-Mills theory.  The technique is
inspired by recent work on ``deconstruction''. Part of the target
theory's supersymmetry is realized exactly on the lattice, reducing or
eliminating the need for fine tuning.   
(Talk based on the paper {\it Supersymmetry on a Spatial
            Lattice}, hep-lat/0206019, by D.B.K., Emmanuel Katz and
          Mithat Unsal)

\vspace{1pc}
\end{abstract}

\maketitle

\section{Exact supersymmetry on the lattice}

Supersymmetric gauge theories are expected to exhibit various
fascinating phenomena, including
electromagnetic 
duality, nontrivial conformal
fixed points, monopole condensation, and relations to gravity and string
theory through the AdS/CFT correspondence. It is
desirable to study these theories 
nonperturbatively, and the lattice is the obvious tool.   There has been much work done on
lattice supersymmetry, but the prospects for practical success presently seem
limited. 

The origin of the problem is that supersymmetry is part of the
super-Poincar\'e group,  which is explicitly broken by the lattice. 
Ordinary Poincar\'e invariance is also broken in lattice QCD, for
example, but  due
to the hypercubic symmetry, operators which violate Poincar\'e
symmetry are
all irrelevant. In a supersymmetric theory, however, the lattice point
group is never sufficient  to forbid all relevant or marginal supersymmetry violating
operators.  The most benign four dimensional supersymmetric gauge
theory is $N=1$ super Yang-Mills (SYM), consisting of gauge bosons and gauginos.  In this
theory, the only relevant SUSY violating operator is the gaugino
mass.  One can therefore tune to the massless point (which has an
enhanced $Z_{2N_c}$ symmetry), or use chirally improved fermions
such as domain wall or overlap fermions to eliminate the fine-tuning
 (see \cite{Montvay:2001aj} and references therein). 

However, most supersymmetric theories contain scalar bosons, such as
  $N=2$ or 
$N=4$ SYM, or $N=1$ theories with matter fields.  When
scalars are present there are typically a plethora of possible SUSY
violating relevant operators to fine tune away, a practically
impossible task.  The only symmetry that can prevent these unwanted
operators is supersymmetry itself.  It is therefore natural to ask
whether one can construct lattices which realize exactly at least {\it some}
of the target theory's supersymmetry, with the hope of ameliorating
the fine tuning problem.  In this talk I describe a recent attempt
along this line \cite{Kaplan:2002wv} (for other recent ideas about
lattice supersymmetry, see \cite{Catterall:2001wx,Fujikawa:2002ic}).
 Presented
here are spatial lattices in Minkowski 
time;
construction of the more useful Euclidean spacetime lattices is underway.

\section{Lattices from orbifolding}

Motivated by recent work on deconstruction of supersymmetric
theories \cite{Arkani-Hamed:2001ca,Arkani-Hamed:2001ie} (see also
\cite{Rothstein:2001tu}), we construct our spatial lattices 
by ``orbifolding''.  We start with
a non-latticized ``mother theory'' which exists in $0+1$ dimensions,
which is a $U(k N^d)$ gauge theory possessing the amount of supersymmetry desired
of the target theory.   It will also possess
a global symmetry (called an $R$-symmetry) which does not commute with
supersymmetry.  To create the lattice we now identify a
$Z_N^d$ subgroup embedded within  both the gauge and global symmetry
groups of the mother theory; the embedding is
discussed in ref. \cite{Kaplan:2002wv}.  We now project out all field
components in the mother theory
which are not singlets under this $Z_N^d$ symmetry.  After projection, adjoint fields of
the mother theory, written as $k N^d\times k N^d$ matrices,  are zero
everywhere except for $N^d$ $k\times k$ blocks on or near the diagonal.
These can be interpreted as fields with near-neighbor interactions
living on a $d$-dimensional spatial lattice with $N^d$ sites.  The
orbifold projection breaks the 
gauge symmetry down to $U(k)^{N^d}$, appropriate for a 
 $U(k)$ gauge theory in the continuum.  The
orbifold projection also breaks some of the mother theory's
supersymmetries, half for each $Z_N$ factor. The
various components of the mother theory's
supermultiplets  get spread about the lattice
within approximately one lattice spacing of each other, so that the breaking of the
mother theory's supersymmetry is rather benign, and is restored in the
continuum limit.

The resulting lattices are quite peculiar and wonderful: there are one-component
fermions scattered over links and sites; noncompact
link variables that become gauge fields;   bosonic
variables on links transforming non-trivially under the
lattice point symmetry which become spin-zero particles in
the continuum.  There is no fermion doubling problem (the
exact residual supersymmetry of the lattice ensures that) and yet one
can realize without fine tuning continuum target theories which exhibit
chiral symmetries.  

\section{A $1+1$ dimensional  example}

As an example of the type of lattice that results, consider our
simplest case: a spatial lattice whose target theory is $(2,2)$ SYM in
$1+1$ dimensions. (The ``$(2,2)$'' designation means that there are
four real chiral supercharges, two left- and two right-handed)  The target
theory consists of a $U(k)$ gauge field with coupling $g_2$, a Dirac fermion $\Psi$, and a
complex scalar $S$ (all $U(k)$ adjoints) with the Lagrangian
\beq
\CL \myeq  \frac{1}{g_2^2}\Tr\left(-\frac{1}{4} F_{\mu\nu}F^{\mu\nu}
 -\mybar\Psi 
  i\sla{D}\Psi - (D_\mu S)^\dagger (D^\mu S) \right.\cr  && + \left.\sqrt{2}(\mybar \Psi_L
  [ S,\Psi_R] + h.c.) - 
 \frac{1}{2}[  S^\dagger, S\,]^2\right)
\eqn{targ}
\eeq
To obtain the lattice theory we start with a $0+1$ dimensional mother theory with a
$U(kN)$ gauge symmetry and four exact supercharges.  This mother
theory is easily constructed by taking the familiar $N=1$ SYM theory
in $3+1$ dimensions, and dimensionally reducing it to $0+1$ dimensions
(that just means: take the gauge and
gaugino fields to be independent of  the $x$, $y$, $z$
coordinates).  This mother theory then has four real bosonic fields
and four real fermionic fields.  It possesses  a global $R$ symmetry which is
$U(1)\times SO(3)$; the $U(1)$ is the same $U(1)$ $R$-symmetry found
in the $3+1$ dimensional $N=1$ SYM theory, while the $SO(3)$ is just the
rotational group that is inherited as an internal symmetry after the
dimensional reduction.  

We now orbifold the theory, modding out by a
$Z_N$ symmetry which is embedded in this $U(kN)\times U(1)\times
SO(3)$ symmetry of the mother theory, as described  in
ref. \cite{Kaplan:2002wv}.  The effect of the orbifolding is to create
an $N$-site, one dimensional periodic lattice. At each site
lives a gauge field $v_0$, a real scalar $\sigma$ and a complex
one-component fermion $\lambda$.  On each link lives a complex scalar
field $\phi$ and a complex one-component fermion $\psi$. Each field is
a $k\times k$ matrix, and there is an independent $U(k)$ gauge
symmetry associated with each site.  Orbifolding breaks
up the multiplet of the mother theory, and reduces the exact
supersymmetry from four supercharges to two. Under the residual
supersymmetry, the site fields form a real (``vector'')
supermultiplet, while the link fields form a ``chiral''
supermultiplet. In component form, the lattice Lagrangian takes the
form:
\beq
L \myeq \frac{1}{g^2}\sum_n \Tr \Bigl[\frac{1}{2}(D_0\sigma_n)^2+
  \,\mybar{\lambda}_n \, iD_0 \,\lambda_n
 + |D_0\, \phi_n|^2 \cr \myplus  \mybar{\psi}_n\, i D_0\psi_n \Bigr.
- \mybar{\lambda}_n[\sigma_n , \lambda_n] +
 \mybar{\psi}_n(\sigma_n  \psi_n-\psi_n\sigma_{n+1})
\cr
\myminus
\left| \sigma_n\phi_n-\phi_n\sigma_{n+1}\right|^2
-\frac{1}{2}\left(\phi_n\mybar\phi_n-\mybar\phi_{n+1}\phi_{n+1}\right)^2
\cr
\myminus
\left. \sqrt{2}\left(i\mybar\phi_n(\lambda_n\psi_n+\psi_n\lambda_{n+1}) +
   h.c.\right)\right]
\ ,  
\eqn{tcomp}
\eeq
where
\beq
D_0\phi_n = \partial_0\phi_n + iv_{0,n}\phi_n -i\phi_nv_{0,n+1}
\eeq
and similarly for $D_0\psi_n$. However, this form hides the
supersymmetry of the lattice.  If instead one writes the Lagrangian in
terms of the appropriate superfields, one finds the simple form
\beq
L = \frac{1}{ g^2} \sum_{n=1}^N\Tr\left[\frac{1}{2}\mybar\Phi_n i \CD^-_0 
\Phi_n + \frac{1}{8}\mybar\Upsilon_n \Upsilon_n 
\right]_{\mybar\theta\theta}\ .
\eeq
where $\Upsilon_n$ is the Grassman chiral multiplet containing the gauge
kinetic terms at site $n$.

This Lagrangian has a classical moduli space (that is: noncompact flat directions
for boson field vevs).  We now expand about the vev 
$\phi_n =   \frac{1}{\sqrt{2a}}\times {\bf 1}_k\ ,
$
where $ {\bf 1}_k$ is the $k\times k$ unit matrix and $a$ will be the
scale defining the lattice spacing.  If we define the target
theory's coupling $g_2$ in terms of the lattice coupling $g$ as $
g_2^2\equiv a g^2$ and lattice size $L=N a$,
then  the continuum limit is $a\to 0$
and $N\to \infty$ for fixed $L^2 g_2^2$.
At tree level we recover the target theory of \Eq{targ}, with the identification
$A_0 = v_0$ and $A_1={\rm Im} \phi$ (where $A_{0,1}$ are the gauge
fields of the target theory) and 
\beq
S=\frac{\sigma + i {\rm Re} \phi}{\sqrt{2}}\
,\ \Psi=\pmatrix{\psi\cr \mybar\lambda}\ .
\eeq

Dimensional analysis reveals that the only dangerous relevant or
marginal operators are scalar tadpoles and mass terms; however the two
exact supersymmetries preclude generating local counterterms for these
operators.   One must, however, control the noncompact flat
directions, or else the path integral is not defined.  This can be
done by fixing initial and final data  on all bosonic zeromodes, and
by restricting correlation measurements to  time separation no longer than the
spatial size of the lattice.

\section{Other lattices}

We have constructed  spatial lattices corresponding to the continuum
super Yang-Mills theories with four real supercharges in $1+1$
dimensions (described above); eight real supercharges in $1+1$
and $2+1$ dimensions; and sixteen real supercharges in $1+1$,
$2+1$ and $3+1$ dimensions ($N=4$ SYM).  These lattices have unusual
structure, such as hexagonal and body-centered cubic for the sixteen
supercharge lattices.  In each case but the last we
can use the lattice symmetries to argue that in perturbation theory there are no allowed
counterterms for relevant or marginal operators that would violate the symmetries
of the target theory; since these are all super-renormalizable
theories, this argument should suffice for proving that no fine tuning
is needed in the continuum limit.  For the four dimensional case, we
have no symmetry argument to forbid dimension four operators that
would violate the desired $N=4$ supersymmetry.  However, we believe
that an anisotropic lattice, where two spatial dimensions are
characterized by a shorter lattice spacing than the third, can have an
$N=4$ continuum limit without fine tuning, where would-be log
divergences are suppressed by a ratio of lattice spacings, which can
be taken to zero.

Euclidean spacetime lattices may be constructed by
orbifolding certain  supersymmetric matrix models.  It
remains to be seen how much fine tuning, if any,
is needed for these lattices.  An encouraging result, however, is that
the fermion ``determinants'' for these theories are real and positive,
allowing for Monte Carlo simulation without a sign problem.

\bibliography{l2002}

\providecommand{\href}[2]{#2}\begingroup\raggedright\begin{thebibliography}{1}

\bibitem{Montvay:2001aj}
I.~Montvay {\em Int. J. Mod. Phys.} {\bf A17} (2002) 2377--2412,
  [\href{http://xxx.lanl.gov/abs/http://arXiv.org/abs/hep-lat/0112007}{{\tt
  http://arXiv.org/abs/hep-lat/0112007}}].

\bibitem{Kaplan:2002wv}
D.~B. Kaplan, E.~Katz, and M.~Unsal
  \href{http://xxx.lanl.gov/abs/http://arXiv.org/abs/hep-lat/0206019}{{\tt
  http://arXiv.org/abs/hep-lat/0206019}}.

\bibitem{Catterall:2001wx}
S.~Catterall and S.~Karamov {\em Nucl. Phys. Proc. Suppl.} {\bf 106} (2002)
  935--937,
  [\href{http://xxx.lanl.gov/abs/http://arXiv.org/abs/hep-lat/0110071}{{\tt
  http://arXiv.org/abs/hep-lat/0110071}}].

\bibitem{Fujikawa:2002ic}
K.~Fujikawa {\em Nucl. Phys.} {\bf B636} (2002) 80--98,
  [\href{http://xxx.lanl.gov/abs/http://arXiv.org/abs/hep-th/0205095}{{\tt
  http://arXiv.org/abs/hep-th/0205095}}].

\bibitem{Arkani-Hamed:2001ca}
N.~Arkani-Hamed, A.~G. Cohen, and H.~Georgi {\em Phys. Rev. Lett.} {\bf 86}
  (2001) 4757--4761,
  [\href{http://xxx.lanl.gov/abs/http://arXiv.org/abs/hep-th/0104005}{{\tt
  http://arXiv.org/abs/hep-th/0104005}}].

\bibitem{Arkani-Hamed:2001ie}
N.~Arkani-Hamed, A.~G. Cohen, D.~B. Kaplan, A.~Karch, and L.~Motl
  \href{http://xxx.lanl.gov/abs/http://arXiv.org/abs/hep-th/0110146}{{\tt
  http://arXiv.org/abs/hep-th/0110146}}.

\bibitem{Rothstein:2001tu}
I.~Rothstein and W.~Skiba {\em Phys. Rev.} {\bf D65} (2002) 065002,
  [\href{http://xxx.lanl.gov/abs/http://arXiv.org/abs/hep-th/0109175}{{\tt
  http://arXiv.org/abs/hep-th/0109175}}].

\end{thebibliography}\endgroup
\bibliographystyle{JHEP} 
\end{document}